\documentclass[a4paper]{article}
\pdfoutput=1

\usepackage[T1]{fontenc}
\usepackage[utf8]{inputenc}
\usepackage[margin=30mm]{geometry}
\usepackage{lmodern}
\usepackage{amsmath,amssymb}
\usepackage{microtype}

\usepackage{amsthm}
\theoremstyle{plain}
\newtheorem{theorem}{Theorem}[section]
\newtheorem{lemma}[theorem]{Lemma}
\newtheorem{corollary}[theorem]{Corollary}
\theoremstyle{definition}
\newtheorem{definition}[theorem]{Definition}
\theoremstyle{remark}
\newtheorem{example}[theorem]{Example}

\usepackage[colorlinks,linkcolor=blue,urlcolor=blue,citecolor=blue]{hyperref}

\usepackage{pgf}

\usepackage{sectsty}
\allsectionsfont{\sffamily\bfseries}
\title{\sffamily\bfseries Scientific Modelling with Coalgebra--Algebra Homomorphisms}

\author{Baltasar Tranc{\'o}n y Widemann \\[\smallskipamount] \normalsize Ilmenau University of Technology \\ {\ttfamily\normalsize baltasar.trancon@tu-ilmenau.de} \and  Michael Hauhs \\[\smallskipamount] \normalsize University of Bayreuth \\ {\ttfamily\normalsize michael.hauhs@uni-bayreuth.de}}

\date{Draft Revision 1}

\usepackage[cmtip,all]{xy}
\newcommand\Lind[1][]{\mathcal{L}_{\mathrm{#1}}}
\newcommand\cata[1]{(\kern -2pt\lvert #1 \rvert\kern -2pt)}
\newcommand\ana[1]{[\kern -2pt( #1 )\kern -2pt]}
\newcommand\klex[2][]{#2^{\star#1}}
\newcommand\klox[1]{#1^\circ}
\newcommand\tsum{\mathop{\textstyle\sum}}
\newcommand\llist{\langle}
\newcommand\rlist{\rangle}
\newcommand\nil{\llist\,\rlist}
\newcommand\cons{\mathbin{::}}

\begin{document}

\maketitle

\begin{abstract}
  Many recursive functions can be defined elegantly as the unique
  homomorphisms, between two algebras, two coalgebras, or one each,
  that are induced by some universal property of a distinguished
  structure.  Besides the well-known applications in recursive
  functional programming, several basic modes of reasoning about
  scientific models have been demonstrated to admit such an exact
  meta-theory.  Here we explore the potential of coalgebra--algebra
  homomorphism that are not a priori unique, for capturing more
  loosely specifying patterns of scientific modelling.  We investigate
  a pair of dual techniques that leverage (co)monadic structure to
  obtain reasonable genericity even when no universal properties are
  given.  We show the general applicability of the approach by
  discussing a suprisingly broad collection of instances from
  real-world modelling practice.
\end{abstract}

\section{Introduction}
\label{intro}

This paper explores a (co)algebraic framework for homomorphic and
recursive reasoning in and about scientific models; that is,
mathematical structures we think in, loaded with interpretations of
phenomena from the world we live in.  To this end, the paper is
structured as follows: the remainder of section~\ref{intro} reviews
the relevant concepts and notations of categorial (co)algebra, with an
interlude in section~\ref{vision} that states the motivation and goal
more precisely, once terms have been established.
Section~\ref{kleisli} introduces our dual pair of recursion schemes of
interest.  Sections~\ref{app} and \ref{coapp} discuss two application
domains per scheme.  These should be regarded as a set of four related
short papers, and are mostly self-contained with respect to discussion
and related work.  A general conclusion is difficult because of the
wide variety of scope, and has to be omitted due to space constraints.
We trust the application subsections to speak for themselves.  Proofs
and illustrations are relegated to the appendix.

\subsection{(Co)Algebras of a Functor}

Let $F$ be an (endo)functor on a category $C$, in all of the
following, but without loss of generality, the category of sets.
$F$-algebras are structures $(X, f : FX \to X)$, where $X$ is called
the \emph{carrier} and $f$ the \emph{operation}.  $F$-coalgebras are
the dual structures $(X, f : X \to FX)$.  Homomorphisms are morphisms
that make a square commute.  They can be defined between two algebras
or two coalgebras, or between one of each:
\begin{align*}
  \xymatrix{
    FX \ar[r]^{f} \ar[d]_{Fh} & X \ar[d]^{h}
    \\
    FY \ar[r]_{g} & Y
  } &&
  \xymatrix{
    X \ar[r]^{f} \ar[d]_{h} & FX \ar[d]^{Fh}
    \\
    Y \ar[r]_{g} & FY
  } &&
  \xymatrix {
    X \ar[d]_{h} \ar[r]^{f} & FX \ar[d]^{Fh}
    \\
    Y & FY \ar[l]^{g}
  } &&
  \xymatrix {
    FX \ar[d]_{Fh} \ar[r]^{f} & X \ar[d]^{h}
    \\
    FY & Y \ar[l]^{g}
  }
\end{align*}
The former two cases of pure (co)algebra homomorphisms have been
studied quite extensively in the frameworks of universal (co)algebra.
The latter two mixed cases have only relatively recently received
significant attention.  The last case, of homomorphisms from an
algebra to a coalgebra, remains rather obscure.  By contrast,
homomorphisms from a coalgebra to an algebra, called
coalgebra--algebra homomorphisms and in the following abbreviated as
\emph{ca-homomorphisms}, are a very general and expressive framework
for divide\&conquer schemes of recursion.  The common intuition
(paraphrased from \cite{Capretta2004}) is the following: `In order to
solve ($h$) a computation problem on complex inputs ($X$), decompose
($f$) them into a collection of subproblems ($FX$), solve these
independently ($Fh$) to obtain a collection of subresults ($FY$), and
compose ($g$) those to form the final result ($Y$).'

The composition of two $F$-(co)algebra homomorphisms is again an
$F$-(co)algebra homomorphism, respectively.  The composition of an
$F$-algebra homomorphism with an $F$-ca-homomorphism, or dually an
$F$-ca-homomorphism with an $F$-coalgebra homomorphism, is again an
$F$-ca-homomorphism.
The following proposition about ``new ca-homomorphisms from old ones''
is less obvious, but useful for the subsequent discussion.

\begin{lemma}
  \label{lemma:alg-tup}%
  Let functor $F$ preserve (co)products.  Then the (co)tuplings of
  ca-homomor\-phisms are the ca-homomorphisms associated with
  (co)product-structured (co)algebras, respectively, up to unique
  isomorphism.
\end{lemma}

\subsection{Distinguished (Co)Algebras}

The $F$-algebras and $F$-algebra homomorphisms form a category
$\mathbf{Alg}(F)$.  Of particular interest are \emph{initial}
$F$-algebras $(\boldsymbol\mu F, \mathrm{in})$.\footnote{Note the
  boldface typography that distinguishes fixpoint operator
  $\boldsymbol\mu$ and monad multplication $\mu$.}  For any
$F$-algebra $(X, f)$ there is a unique homomorphism $\cata{f} :
(\boldsymbol\mu F, \mathrm{in}) \to (X, f)$, called a
\emph{catamorphism}.  By Lambek's lemma, the operation ($\mathrm{in}$)
is a bijection.
Dually, the $F$-coalgebras and $F$-coalgebra homomorphisms form a
category $\mathbf{Coalg}(F)$.  Of particular interest are \emph{final}
$F$-coalgebras $(\boldsymbol\nu F, \mathrm{out})$.  For any
$F$-coalgebra $(X, f)$ there is a unique homomorphism $\ana{f} : (X,
f) \to (\boldsymbol\nu F, \mathrm{out})$, called an
\emph{anamorphism}.  The operation ($\mathrm{out}$) is a bijection.
For initial algebras and final coalgebras of a variety of well-behaved
functors, see \cite{Rutten2000}.

For ca-homomorphisms, the situation is more complex, because they do
not compose among themselves.  A $F$-algebra $(Y, g)$ is called
\emph{corecursive}~\cite{Capretta2009} if and only if it plays a role
analogous to a final $F$-coalgebra: For any $F$-coalgebra $(X, f)$
there is a unique $F$-ca-homomorphism $h : (X, f) \to (Y, g)$.
Dually, a $F$-coalgebra $(X, f)$ is called
\emph{recursive}~\cite{Capretta2004,Osius1974} if and only if it plays
a role analogous to an initial $F$-algebra: For any $F$-algebra $(Y,
g)$ there is a unique $F$-ca-homomorphism $h : (X, f) \to (Y, g)$.
Trivial examples are obtained by Lambek's lemma: $(\boldsymbol\mu F,
\mathrm{in}^{-1})$ is a recursive $F$-coalgebra; and dually
$(\boldsymbol\nu F, \mathrm{out}^{-1})$ is a corecursive $F$-algebra.

\subsection{Interlude: Vision}
\label{vision}

The present paper deals with intermediate situations: Neither are our
ca-homomorphisms of interest completely arbitrary, with algebra and
coalgebra chosen ad hoc at the same level of particularity; nor is
there an obvious candidate for a general (co)recursive (co)algebra,
leaving only the partner to be chosen in particular, and the
homomorphism induced uniquely.  Instead, we investigate situations
where either the algebra or the coalgebra part is more central, and
the possibility and general shape of homomorphisms is to be studied,
independently from its less fixed partner.

This line of investigation ties into our overarching research
programme on the recursive nature of scientific modelling posited by
Rosen~\cite{Rosen1991}.  We shall identify certain modes of abstract
formal reasoning over classes of models as (co)algebras, and their
rigorous interpretation as ``reusable'' ca-homomorphisms between a
fixed general formal language and a variety of particular models.

In previous work, we have investigated dual, purely algebraic and
coalgebraic reasoning modes, and identified them as `queries of
causality' and `representations of behavior', respectively
\cite{Hauhs2010a}.  We have also explored various types of mixed
modes, and drawn connections between structural operational semantics
and cellular automata \cite{TranconyWidemann2011}, and between
course-of-value iteration and history-dependent dynamics
\cite{TranconyWidemann2014a}, respectively.  The present work deals
with yet more types of mixed mode, but ones that come as a dual pair,
as in \cite{Hauhs2010a}.  We consider the duality to be of great
philosophical interest, although the present modest paper can only
give a few first directions.

A great deal of the desired reusability shall be achieved by imposing
a (co)monad structure on the functor $F$.  In the following section,
we review the relevant concepts and notations very briefly, aware that
our presentation can only serve as a glossary, not an introduction.

\subsection{(Co)Monads}

A \emph{monad} $\mathbb T = (T, \eta, \mu)$ is an endofunctor $T$
together with two natural transformations $\eta : 1 \Rightarrow T$ and
$\mu : TT \Rightarrow T$, called \emph{unit} and
\emph{multiplication}, respectively, such that $\mu \circ \eta T =
\mathrm{id} T = \mu \circ T \eta$ and $\mu \circ \mu T = \mu \circ T
\mu$.  A monad gives rise to a \emph{Kleisli extension} operator that
sends any morphism $f : X \to TY$ to a unique $\klex{f} : TX \to TY$,
such that $\klex{\eta} = \mathrm{id}T$, $\klex{f} \circ \eta_X = f$
and $\klex{g} \circ \klex{f} = \klex{(\klex{g} \circ f)}$; namely
$\klex{f} = \mu_Y \circ Tf$ and $f = \klex{f} \circ \eta_X$.

The \emph{Kleisli category} $\mathbf{Kl}(\mathbb T)$ of a monad has
the same objects as the underlying category, but homomorphisms $f : X
\xrightarrow{\mathbf{Kl}(\mathbb T)} Y$ whenever $f : X \to TY$.  Its
identity is $\mathrm{id}^{\mathbf{Kl}(\mathbb T)} = \eta$; composition
is given by $g \circ^{\mathbf{Kl}(\mathbb T)} f = \klex{g} \circ f$.
Coalgebras of a monadic functor are of Kleisli type $f : X
\xrightarrow{\mathbf{Kl}(\mathbb T)} X$ and can be iterated.  We write
$\klex[,n]{f}$ for the morphism such that $\klex[,n]{f} =
(\klex{f})^n$.

For many nice endofunctors $F$, a \emph{free} monad $\mathbb F^*$ can
be constructed as follows~\cite{Barr1970},
\begin{align*}
  F^*X &= \boldsymbol\mu (X + F) &
  F^*(f : X \to Y) &= \cata{\mathrm{in}_Y \circ (f + \mathrm{id}_{FF^*Y})}
\end{align*}
$F^*$ is a functor, and turns the family of initial algebras into a
natural transformation $\mathrm{in} : \mathrm{Id} + FF^* \Rightarrow
F^*$.  Besides the usual natural transformations that go with a monad,
define two additional ones, $\tau : FF^* \Rightarrow F^*$ and $\kappa
: F \Rightarrow F^*$ (adapted from~\cite{Bonchi2014}), as well as
Kleisli extension, simultaneously as:
\begin{align*}
  \eta &= \mathrm{in} \circ \iota_1 &
  \tau &= \mathrm{in} \circ \iota_2 &
  \mu &= \klex{(\mathrm{id}F^*)} &
  \kappa &= \tau \circ F \eta &
  \klex{(f : X \to F^*Y)} &= \cata{[f, \tau_Y]}
\end{align*}

\begin{lemma}
  \label{lemma:tau}%
  The extra transformations obey the law $\tau = \mu \circ \kappa F^*$.
\end{lemma}

Dually, a \emph{comonad} is an endofunctor $D$ together with two
natural transformations $\varepsilon : D \Rightarrow 1$ and $\nu : D
\Rightarrow DD$, called \emph{counit} and \emph{comultiplication},
respectively, such that $\varepsilon D \circ \nu = \mathrm{id} D = D
\varepsilon \circ \nu$ and $\nu D \circ \nu = D \nu \circ \nu$.  A
comonad gives rise to a \emph{co-Kleisli extension} operator that
sends any morphism $f : DX \to Y$ to a unique $\klox{f} : DX \to DY$,
such that $\klox{\varepsilon} = \mathrm{id}D$, $\varepsilon_Y \circ
\klox{f} = f$ and $\klox{g} \circ \klox{f} = \klox{(g \circ
  \klox{f})}$; namely $\klox{f} = Df \circ \nu_X$ and $f =
\varepsilon_Y \circ \klox{f}$.

The \emph{co-Kleisli category} $\mathbf{Cl}(\mathbb D)$ of a comonad
has the same objects as the underlying category, but homomorphisms $f
: X \stackrel{\mathbf{Cl}(\mathbb D)}\to Y$ whenever $f : DX \to Y$.
Its identity is $\mathrm{id}^{\mathbf{Cl}(\mathbb D)} = \varepsilon$;
composition is given by $g \circ^{\mathbf{Cl}(\mathbb D)} f = g \circ
\klox{f}$.

A \emph{cofree} comonad $F^\infty$ can often be constructed
from an endofunctor $F$ as follows.
\begin{align*}
  F^\infty X &= \boldsymbol\nu (X \times F) &
  F^\infty(f : F^\infty X \to Y) &= \ana{(f \times \mathrm{id}_{FF^\infty X}) \circ \mathrm{out}_X}
\end{align*}
$F^\infty$ is a functor, and turns the family of final coalgebras into a
natural transformation $\mathrm{out} : F^\infty \Rightarrow \mathrm{Id} +
FF^\infty$.  Besides the usual natural transformations that go with a
comonad,
define two additional ones, $\theta : F^\infty \Rightarrow FF^\infty$
and $\chi : F^\infty \Rightarrow F$ (dually extrapolated
from~\cite{Bonchi2014}), as well as co-Kleisli extension,
simultaneously as:
\begin{align*}
  \varepsilon &= \pi_1 \circ \mathrm{out} &
  \theta &= \pi_2 \circ \mathrm{out} &
  \nu &= \klox{(\mathrm{id}F^\infty)} &
  \chi &= F \varepsilon \circ \theta &
  \klox{(f : F^\infty X \to Y)} &= \ana{\langle f, \theta_X \rangle}
\end{align*}

\begin{lemma}
  \label{lemma:theta}%
  The extra transformations obey the law $\theta = \chi
  F^\infty \circ \nu$.
\end{lemma}

\subsection{(Co)Algebras of a (Co)Monad}

An algebra $(X, f)$ of a functor $T$ is also an algebra of the monad
$\mathbb T = (T, \eta, \mu)$, also called an Eilenberg--Moore
algebra, if and only if $f$ is compatible with the monad operations,
that is $f \circ \eta_X = \mathrm{id}_X$ and $f \circ \mu_X = f \circ
Tf$.
The algebras of a monad $\mathbf{Alg}(\mathbb T)$ form a full
subcategory of the algebras $\mathbf{Alg}(T)$ of the underlying
functor.
The monad multiplication is a distinguished algebra operation:

\begin{lemma}
  \label{lemma:mu}%
  The $\mathbb T$-algebras of the form $(TX, \mu_X)$ are ``locally''
  weakly initial: for every $\mathbb T$-algebra $(X, f)$, there is a
  canonical $T$-algebra homomorphism $f : (TX, \mu_X) \to (X, f)$.
  Furthermore, any morphism $h : X \to Y$ gives rise to a $T$-algebra
  homomorphism $Th : (TX, \mu_X) \to (TY, \mu_Y)$.
\end{lemma}

The algebras of a functor $F$ and the algebras of the free monad
$\mathbb F^* = (F^*, \eta, \mu)$ are in one-to-one correspondence:
Each $\mathbb F^*$-algebra $(X, f)$ is of the form $f = e^\sharp =
\cata{[\mathrm{id}_X, e]}$, generated uniquely by the $F$-algebra $(X,
e)$ where $e = f^\flat = f \circ \kappa_X$.  Thus we can reformulate
Lemma~\ref{lemma:tau}.

\begin{lemma}
  \label{lemma:tau'}%
  The free monad extra transformations obey the law $\tau = \mu^\flat$.
\end{lemma}

Dually, a coalgebra $(X, f)$ of a functor $D$ is also a coalgebra of
the comonad $\mathbb D = (D, \varepsilon, \nu)$, if and only if
$\varepsilon_X \circ f = \mathrm{id}_X$ and $\nu_X \circ f = Df \circ f$.
The coalgebras of a comonad $\mathbf{Coalg}(\mathbb D)$ form
a full subcategory of the coalgebras $\mathbf{Coalg}(D)$ of the
underlying functor.

\begin{lemma}
  \label{lemma:nu}%
  The $\mathbb D$-coalgebras of the form $(D X, \nu_X)$ are
  ``locally'' weakly final: for every $\mathbb D$-coalgebra $(X, f)$,
  there is the canonical $D$-coalgebra homomorphism $f : (X, f) \to (D
  X, \nu_X)$.  Furthermore, any morphism $h : X \to Y$ gives rise to a
  $D$-coalgebra homomorphism $Dh : (DX, \nu_X) \to (DY, \nu_Y)$.
\end{lemma}

The coalgebras of a functor $F$ and the coalgebras of the cofree
comonad $\mathbb F^\infty = (F^\infty, \varepsilon, \nu)$ are in one-to-one
correspondence: Each $\mathbb F^\infty$-coalgebra $(X, f)$
is of the form $f = e^\sharp = \ana{\langle\mathrm{id}_X, e\rangle}$,
generated uniquely by the $F$-coalgebra $(X, e)$ where $e = f^\flat =
\chi_X \circ f$.

\begin{lemma}
  \label{lemma:theta'}%
  The cofree comonad extra transformations obey the law $\theta = \nu^\flat$.
\end{lemma}

\section{(Co)Kleisli (Co)Induction}
\label{kleisli}

We shall demonstrate that well-behaved classes of interesting
ca-homomorphisms arise from a pair of dual recursion schemes that
leverage (co)monadic structure.  The monadic version has been studied
recently in the context of trace semantics \cite{Bonchi2014}.
Although its dual comonadic twin is straightforward, we are not aware
of previous uses; in fact we shall argue why its characteristic
properties, albeit essentially useful for our purpose here, are
undesirable in many circumstances.

\subsection{Kleisli Coinduction}

\begin{definition}[Kleisli Coinduction]
  Let $\mathbb T = (T, \eta, \mu)$ be a monad.  Let $(X, e)$ be a
  $T$-coalgebra.  A morphism $e^\dag : X \to TY$ is called a
  \emph{Kleisli-coinductive solution} of \emph{equation} $e$, if and
  only if $e^\dag = e^\dag \circ^{\mathbf{Kl}(\mathbb T)} e$.
\end{definition}

\begin{lemma}[Characterization]
  \label{lemma:dag}%
  Kleisli-coinductive solutions $e^\dag : X \to TY$ are precisely the
  $T$-ca-homomor\-phisms to the multiplicative $\mathbb T$-algebra
  $(TY, \mu_Y)$.
\end{lemma}

A Kleisli-coinductive solution $e^\dag : X \to TY$ of $e : X \to TX$
appears as a morphism of type $X \to Y$ in the Kleisli category.  As
usual for corecursive functions, its codomain $Y$ is not determined by
$e$ at all.

\begin{lemma}[Substitution]
  \label{lemma:dag-subst}%
  Let $(X, e)$ be a $T$-coalgebra and $h : Y \to Z$ be any morphism.
  If $e^\dag : X \to TY$ is a Kleisli-coinductive solution, then so is
  $Th \circ e^\dag : X \to TZ$.
\end{lemma}

Our motivation for studying this particular corecursion scheme is that
each Kleisli-coinductive solution extends to many ca-homomorphisms in
a regular way, such that its common properties can be studied
regardless of the target algebra.

\begin{theorem}[Universality]
  \label{theorem:dag}%
  Given \emph{any} $T$-coalgebra $(X, e)$ and Kleisli-coinductive
  solution $e^\dag : X \to TY$, there is a canonical family of
  extensions to $T$-ca-homomorphisms $(f \circ e^\dag)$ into \emph{all}
  $\mathbb T$-algebras $(Y, f)$.
\end{theorem}

\begin{corollary}[Recursivity]
  \label{corollary:rec}%
  If a $T$-coalgebra $(X, e)$ is recursive, then it has a unique
  Kleisli-coinductive solution, whose extension is the unique
  ca-homomorphism.
\end{corollary}

Conversely, if a $T$-coalgebra has a unique Kleisli-coinductive
solution, then it is ``morally recursive'': there is a unique
ca-homomorphism into any $\mathbb T$-algebra that is canonical in the
sense of Lemma~\ref{lemma:dag}.  Note that $T$-algebras which are
incompatible with the monadic structure are generally not covered.

On the other hand, Kleisli-coinductive solution can be non-unique, or
fail to exist at all.  Hence one should regard the equation coalgebra
as a \emph{specification} rather than \emph{definition} of
ca-homomorphisms.  Such a specification can be contradictory or loose;
the following proposition shows that it may also be tautological.

\begin{lemma}[Unit]
  \label{lemma:tauto}%
  Let $\mathbb T = (T, \eta, \mu)$ be a monad.  For a $T$-coalgebra of
  the form $(X, \eta_X)$, \emph{all} morphisms of type $X \to
  TY$ are Kleisli-coinductive solutions.
\end{lemma}

While a globally tautological specification is evidently of little
practical value, we shall make good use of pointwise local tautology
in the application section~\ref{dyn}.

For the special case of a free monad $\mathbb F^*$, we have an
asymmetric situation: The $F^*$-algebras have a distinguished subclass
of interest, namely the $\mathbb F^*$-algebras which are equivalent to
the simpler class of $F$-algebras.  But $F^*$-coalgebras generally
obey no such constraint of expressivity.  If one is imposed
deliberately, then a reduced characterization of Kleisli-coinductive
solutions in terms of $F$-ca-homomorphisms can be given.

\begin{definition}[Basic Equation]
  Let $\mathbb F^* = (F^*, \eta, \mu)$ be a free monad over
  endofunctor $F$.  An $F^*$-coalgebra $(X, e)$ is called
  \emph{basic}, if and only if it factors as $e = \kappa_X \circ e_0$.
  Note that $(X, e_0)$ is an $F$-coalgebra.
\end{definition}

\begin{lemma}[Demonadization]
  \label{lemma:basic}%
  $F^*$-ca-homomorphisms from basic $F^*$-coalgebras $(X, \kappa_X
  \circ e_0)$ to $\mathbb F^*$-algebras $(Y, f)$ are precisely the
  $F$-ca-homomorphisms from $(X, e_0)$ to $(Y, f^\flat)$.
\end{lemma}

\begin{corollary}
  \label{corollary:basic}%
  The Kleisli-coinductive solutions $e^\dag : X \to F^*Y$ of basic
  $F^*$-coalgebras $(X, \kappa_X \circ e_0)$ are precisely the
  $F$-ca-homomorphisms from $(X, e_0)$ into the $F$-algebra $(F^*Y,
  \tau_Y)$.
\end{corollary}

\subsection{Co-Kleisli Induction}

All of the preceding statements can be dualized.

\begin{definition}[Co-Kleisli Induction]
  Let $\mathbb D = (D, \varepsilon, \nu)$ be a comonad.  Let $(Y, k)$
  be a $D$-algebra.  A morphism $k^\ddag : DX \to Y$ is called a
  \emph{co-Kleisli-inductive solution} of \emph{coequation} $k$, if and
  only if $k^\ddag = k \circ^{\mathbf{Cl}(\mathbb D)} k^\ddag$.
\end{definition}

\begin{lemma}[Characterization]
  \label{lemma:codag}
  Co-Kleisli-inductive solutions $k^\ddag : DX \to Y$ are precisely
  the ca-homomor\-phisms from the $D$-coalgebra $(DX, \nu_X)$.
\end{lemma}

\begin{lemma}[Substitution]
  \label{lemma:codag-subst}
  Let $(Y, k)$ be a $D$-algebra and $h : W \to X$ be any morphism.
  If $k^\ddag : DX \to Y$ is a co-Kleisli-inductive solution, then so is
  $k^\ddag \circ Dh : DW \to Y$.
\end{lemma}

\begin{theorem}[Universality]
  \label{theorem:codag}
  Given \emph{any} $D$-algebra $(Y, k)$ and co-Kleisli-inductive
  solution $k^\ddag : DX \to Y$, there are canonical extensions to
  ca-homomorphisms $(k^\ddag \circ f)$ from \emph{all} $\mathbb
  D$-coalgebras $(X, f)$.
\end{theorem}

\begin{corollary}[Corecursivity]
  \label{corollary:corec}%
  If a $D$-algebra $(Y, k)$ is corecursive, then it has a unique
  co-Kleisli-inductive solution, whose extension is the unique
  ca-homomorphism.
\end{corollary}

\begin{lemma}[Counit]
  \label{lemma:counit}
  Let $\mathbb D = (D, \varepsilon, \nu)$ be a comonad.  For a
  $D$-algebra of the form $(Y, \varepsilon_Y)$, \emph{all} morphisms
  of type $DX \to Y$ are co-Kleisli-inductive solutions.
\end{lemma}

\begin{definition}[Cobasic Coequation]
  Let $\mathbb F^\infty = (F^\infty, \varepsilon, \nu)$ be a cofree
  comonad over endofunctor $F$.  An $F^\infty$-algebra $(Y, k)$ is
  called {\bfseries cobasic}, if and only if it factors as $k = k_0
  \circ \chi_Y$.  Note that $(Y, k_0)$ is an $F$-algebra.
\end{definition}

\begin{lemma}[Decomonadization]
  \label{lemma:cobasic}%
  $F^\infty$-ca-homomorphisms from $F^\infty$-coalgebras $(X, f)$ to
  cobasic $\mathbb F^\infty$-algebras $(Y, k_0 \circ \chi_Y)$ are
  precisely the $F$-ca-homomorphisms from $(X, f^\flat)$ to $(Y,
  k_0)$.
\end{lemma}

\begin{corollary}
  \label{corollary:cobasic}%
  Co-Kleisli-inductive solutions $k^\ddag : F^\infty X \to Y$ of
  cobasic $F^\infty$-algebras are precisely the $F$-ca-homomorphisms
  from the $F$-coalgebra $(F^\infty X, \theta_X)$.
\end{corollary}

Our examples in section~\ref{coapp} build on cofree comonads, which
can be understood to create spaces of non-well-founded node-labelled
trees.  In this context, \emph{induction} seems like a thing out of
hell: co-Kleisli-inductive solutions are shamelessly allowed to depend
on infinite regresses, as well as in vicious circles (via
$\varepsilon$) on their own results.  But, as we have pointed out, we
are happy to sacrifice universal existence and uniqueness of
solutions; the scheme is still quite useful as a formal framework for
organizating \emph{existing} modes of reasoning, whose potential
unsoundness is either of no practical concern, or resolved in
domain-specific ways.

\section{Applications of Kleisli Coinduction}
\label{app}

\subsection{Dynamical Systems}
\label{dyn}

Choose a monoid, written additively as $(\Delta, 0, {+})$.  It may or
may not be Abelian, or a group.  The (left) product with a monoid, $T
= \Delta \times \mathrm{Id}$, is a monad with:
\begin{align*}
  \eta(x) &= (0, x) & \mu\bigl(t, (u,
  x)\bigr) &= (t + u, x)
  & \klex{f}(t, x) &= (t + u, y) \quad\text{where}~ f(x) = (u, y)
\end{align*}
The algebras of this monad are exactly the (left) monoid actions; $(X, f :
\Delta \times X \to X)$ where:
\begin{align*}
  f(0, x) &= x & f(t + u, x) &= f\bigl(t, f(u, x)\bigr)
\end{align*}

Monoid actions are the most general formal framework for dynamical
systems.  The monoid is understood as the structure of durations of
time.  Default candidates are non-negative/all integers/reals, which
model irreversible/reversible discrete/continuous time, respectively.
The language that arises from $T$ deals exclusively with the passing
of time: In covariant positions, such as the codomain of morphisms in
the Kleisli category, $(t, x) \in TX$ reads as ``after $t$~time units,
$x$''; in contravariant positions, such as the domain of $T$-algebras,
$(t, x) \in TX$ reads as ``$t$~time units after $x$''.  The Kleisli
category is the category of sets and timed functions, where $f(x) =
(t, y)$ consequently reads as ``$f$ maps $x$ after $t$~time units
to~$y$''.  Kleisli composition is sequential in time; delays are
accumulated.

\begin{example}[Harmonic Oscillator]
  \label{example:osci}%
  Consider the textbook example of a harmonic oscillator, a point mass
  $m$ moving frictionlessly along a straight line, acted on by a
  restoring force proportional, with coefficient $k$, to its
  displacement $x$.  In a Newton-style modelling approach, this system
  is specified by a second-order linear differential equation $\ddot x
  + \frac km x = 0$, which simply states that all acceleration is due
  to the restoring force.
  
  The appropriate analytic model is a dynamical system with $\Delta
  = \mathbb{R}$ and state vectors $(x, \dot x) \in \mathbb{R}^2$.
  \begin{equation*}
    f\bigl(t, (x, \dot x)\bigr) = \left(\begin{array}{@{}rr@{}}
      \cos \omega t & \omega^{-1} \sin \omega t
      \\
      -\omega \sin \omega t & \cos \omega t
    \end{array}\right)
    \begin{pmatrix}
      x \\ \dot x
    \end{pmatrix}
    \qquad \text{where}~
    \omega = \sqrt{\frac km}
  \end{equation*}
\end{example}

The ca-homomorphism diagram for this monad takes the equation
coalgebra $(X, e)$ as a set $X$ of symbolic states, together with a
map $e$ that gives each state $x \in X$ a backwards-looking
specification $(t, x')$, understood as ``$t$~time units after state
$x'$''.  The ca-homomorphisms $h : (X, e) \to (Y, f)$ are then the
instantiations consistent with the forwards-looking dynamics $f$ on a
concrete state space $Y$.  Note that partial specifications are
included automatically: A state specified tautologically as zero time
units after itself is consistent with any instantiation, by pointwise
Lemma~\ref{lemma:tauto}.

By unfolding the definition, the Kleisli-coinductive solutions of a
$T$-coalgebra $(X, e)$ are exactly the morphisms of type $e^\dag : X
\to TY$ such that $e(x) = (t, x')$ and $e^\dag(x') = (u, y)$ implies
$e^\dag(t + u, y)$.  They have a surprisingly rich structure, and
admit necessary and/or sufficient conditions, under a variety of mild
assumptions concerning the time-likeness of $\Delta$.  As outlined
before, our framework allows us to study the existence and parameters
of solutions, independently of interpretation in a concrete dynamical
system.

\begin{definition}
  \label{def:dyn-rels}%
  Associate with a $T$-coalgebra $(X, e)$ several relations:
  \begin{itemize}
  \item $(\sim_e) \subseteq X \times X$ is the
    reflexive-symmetric-transitive closure of $\pi_2 \circ e$, an
    equivalence.
  \item $(\succeq_e) \subseteq X \times X$ is the reflexive-transitive
    closure of $\pi_2 \circ e$, a preorder.
    \begin{itemize}
    \item Partition into complementary subrelations $(\succ_e) =
      (\succeq_e) \setminus (\preceq_e)$, a strict partial order, and
      $(\simeq_e) = (\succeq_e) \cap (\preceq_e) \subseteq (\sim_e)$,
      an equivalence, as usual.
    \end{itemize}
  \item $(\leadsto_e) \subseteq X \times T X$ is the Kleisli analog of
    $(\succeq_e)$, namely
    $(\leadsto_e) = \bigcup_{n=0}^\infty \klex[,n]{e}$.

    We call $(\leadsto_e)$ {\bfseries consistent}, if and only if $x
    \leadsto_e (t, x)$ implies $t = 0$.
  \end{itemize}
\end{definition}

\noindent These relations allow us to state some basic properties of
Kleisli-coinductive solutions.

\begin{lemma}
  \label{lemma:dyn-coprod}%
  Let $(X, e)$ be a $T$-coalgebra.  Then it splits into a coproduct of
  the $(\sim_e)$-equivalence classes.  Furthermore, every
  Kleisli-coinductive solution $e^\dag : X \to TY$ is locally constant
  in its second component on each $(\sim_e)$-equivalence class: If $x
  \sim_e x'$, then $e^\dag(x) = (u, y)$ and $e^\dag(x') = (u', y')$
  imply $y = y'$.
\end{lemma}

\begin{lemma}
  \label{lemma:dyn-cons}%
  If $(\leadsto_e)$ is consistent, then for each pair $x \succeq_e x'$
  there is a unique $t$ such that $x \leadsto_e (t, x')$.
\end{lemma}

\begin{lemma}
  \label{lemma:dyn-stack}%
  Let $e^\dag : X \to TY$ be a Kleisli-coinductive solution of $(X,
  e)$.  If $x \leadsto_e (t, x')$, and $e^\dag(x') = (u, y)$, then
  $e^\dag(x) = (t + u, y)$.  In particular, $x \leadsto_e (0, x')$
  implies $e^\dag(x) = e^\dag(x')$.
\end{lemma}

We say that the monoid $(\Delta, 0, {+})$ has right cancellation, if
and only if $t + u = u$ implies $t = 0$ for all $t, u \in \Delta$.
For cancellative monoids, a reasonable assumption for standard models
of time, we can give both a necessary and a sufficient condition on
the solvability of a $T$-coalgebra.

\begin{theorem}
  Assume the monoid $(\Delta, 0, {+})$ has right cancellation.  Then a
  $T$-coalgebra $(X, e)$ has Kleisli-coinductive solutions, only if
  $(\leadsto_e)$ is consistent.
\end{theorem}

\begin{proof}
  Assume, for contradiction, that $x \leadsto_e (t, x)$ where $t \neq
  0$.  Let $(u, y) = e^\dag(x)$.  Then also $e^\dag(x) = (t + u, y)$,
  by Lemma~\ref{lemma:dyn-stack}; hence $t + u = u$, and finally by
  cancellation $t = 0$.
\end{proof}

\begin{theorem}
  \label{theorem:dyn-wellf}%
  Assume the monoid $(\Delta, 0, {+})$ has right cancellation.  Then
  a $T$-coalgebra $(X, e)$ has Kleisli-coinductive solutions, if
  $(\leadsto_e)$ is consistent, and $(\succ_e)$ is well-founded.
\end{theorem}

\begin{proof}[Proof (Sketch)]
  By well-founded induction and Lemma~\ref{lemma:dyn-stack}.
\end{proof}

\noindent For reversible time, the conditions can be made tight.

\begin{theorem}
  \label{theorem:dyn-group}%
  Assume $(\Delta, 0, {+})$ is a group.  Then a $T$-coalgebra $(X, e)$
  has Kleisli-coinductive solutions, if (and only if) $(\leadsto_e)$ is
  consistent.
\end{theorem}

\begin{proof}[Proof (Sketch)]
  Componentwise by Lemma~\ref{lemma:dyn-coprod}, then either by
  reduction to Theorem~\ref{theorem:dyn-wellf} or else by inversion of
  Lemma~\ref{lemma:dyn-cons}.
\end{proof}

So far, we have described the solution spaces of $T$-coalgebras
formally, but given no explication of their meaning in the modelling
context.  We shall now demonstrate that the $T$-ca-homomorphisms are a
very abstract and general account of \emph{time series}: discrete
samples of state snapshots over the dynamic evolution of a system.

Fix $\Delta$ as either the nonnegative or all reals.  Fix a
bilaterally infinite sequence of time differences, $(\delta_i \in
\mathbb{R})$ for all $i \in \mathbb{Z}$.  We write $(t_i)$ for the
corresponding partial sums:
\begin{align*}
  t_i &= +\tsum_{j\in[0, i)} \delta_i \quad\text{if}~ i \geq 0 & t_i &=
  -\tsum_{j\in[i,0)} \delta_i \quad\text{if}~ i \leq 0
\end{align*}
Now consider the following $T$-coalgebras on integer intervals:
\begin{align*}
  X_1 &= (-\infty, +\infty) &&& e_1(i + 1) &= (\delta_i, i) && (i \in \mathbb{Z})
  \\
  X_2 &= (-\infty, 0] &&& e_2(i + 1) &= (\delta_i, i) && (i < 0)
  \\
  X_3 &= [0, +\infty) & e_3(0) &= (0, 0) & e_3(i + 1) &= (\delta_i, i) && (0 \leq i)
  \\
  X_4 &= [0, n] & e_4(0) &= (0, 0) & e_4(i + 1) &= (\delta_i, i) && (0 \leq i < n)
\end{align*}
We shall verify that the induced ca-homomorphisms from $e_k$ into
dynamical systems are the different shapes of time series: for $k = 1,
2, 3, 4$, the bilaterally infinite, the left-infinite, the
right-infinite, and the finite of length $n+1$, respectively.

In each case, there is only one equivalence class of $(\sim_{e_k})$.
Hence any Kleisli-coinductive solution $e_k^\dag : X_k \to TY$ fixes a
single target element $y_0 \in Y$, and is of the form
$e_k^\dag(i) = (t_i + u_0, y_0)$.
Consequently, the ca-homomorphisms $h = f \circ e_k^\dag : (X_k, e_k)
\to (Y, f)$ are of the form: $h(i) = f(t_i + u_0, y_0)$.
That is, they are sequences of elements on the trajectory of reference
state $y_0$, spaced according to the sampling time sequence $(t_i)$
relative to reference time $u_0$.  Without loss of generality, $u_0$
can be made to vanish, by rewriting to $h(i) = f(t_i, y_0')$ where
$y_0' = f(u_0, y_0)$.
Furthermore, if $(\delta_i)$ is constant, then we obtain the
\emph{equidistant} time series, with the simpler form $h(i) =
f(i\cdot\delta, y_0)$.

\begin{example}[Period]
  For the harmonic oscillator from Example~\ref{example:osci}, the
  well-known periodic behaviour can be captured very concisely and
  naturally, by stating that the bilaterally infinite, equidistant time
  series specification coalgebra $(X_1, e_1)$ with $\delta =
  2\pi\omega^{-1}$ admits only constant ca-homo\-mor\-phisms.
\end{example}

\begin{example}[Zeno]
  As a sidenote, the pathological class of non-equidistant,
  right-infinite but bounded time series have played an important,
  puzzling role in ancient Greek ``scientific modelling''.  For
  instance, Zeno's stadium run can be specified by $T$-coalgebra
  $(X_3, e_3)$ with $\delta_i = 2^{-(i+1)}$, where half of the
  remaining time is consumed at each step.
\end{example}

This account of discrete time series over continuous dynamical systems
may seem a little contrived, simply because standard textbook
presentations look fairly different, and do not suggest the use of a
formal framework beyond simple set theory and ``index magic''.  The
following example shall serve, among other things, to counter that
impression; there the study of ca-homomorphisms from Kleisli
coinduction leads to observations and structures that are completely
standard in the field.

\subsection{Markov Chains}

Our second example application, although conceptually rather a little
more advanced, can be presented with much less technical detail,
because we can build on standard accounts in terms of far more
expressive mathematics; see for instance \cite{Howard1971}.

Consider discrete distributions on a set $X$, with possibly countably
infinite support, $TX = \bigl\{ \pi : X \to [0, 1] \bigm| \tsum_{x \in
  X} \pi(x) = 1 \bigr\}$.  The distribution space $TX$ has a convex
structure.  Hence its elements can be understood as formal sums
$\tsum_i p_i x_i$, and the operations
\begin{align*}
  Tf(\tsum_i p_i x_i) &= \tsum_i p_i f(x_i) & \eta(x) &= 1x & \mu\bigl(\tsum_i p_i (\tsum_j q_{ij} x_{ij})\bigr) &= \tsum_{ij} p_i q_{ij} x_{ij}
\end{align*}
define a monad $\mathbb T = (T, \eta, \mu)$~\cite{Jacobs2010}.
Alternatively, the elements $\pi \in TX$ can be understood as
stochastic row vectors $\Pi$ with columns index by $X$.  By extension,
Kleisli morphisms of type $f : X \to TY$ can be understood as right
stochastic matrices $F$ of shape $X \times Y$, by currying.

\begin{lemma}
  \label{lemma:markov-matrix}%
  Kleisli extension and composition are right matrix multiplication:
  \begin{align*}
    \klex{f}(\pi) &= \Pi F & g \circ^{\mathbf{Kl}(\mathbb T)} f &= F G
  \end{align*}
\end{lemma}

The $T$-coalgebras $(X, e)$ are exactly the \emph{space-discrete,
  time-homogeneous Markov chains}, a model class pervading almost all
corners of science.  The $\mathbb T$-algebras are \emph{convex sets}:
structures $(Y, f)$ with a set $Y$ closed under a given interpretation
$f$ of formal convex combinations.  In a very abstract but logically
precise sense, the corresponding ca-homomorphisms employ convex sets
as models of the \emph{long-term behaviour} of a Markov chain: Let $Y$
be a set of possible long-term behaviours (refraining from jumping to
the conclusion of well-known candidates).  A ca-homomorphism is a map
$h : X \to Y$ of Markov states to long-term behaviours that is
consistent with transition structure $e$, namely $e(x) = \tsum_i
p_ix_i'$ implies $h(x) = f\bigl(\tsum_i p_i h(x_i')\bigr)$.  That is,
the long-term behaviour associated with a state is an appropriately
weighted convex combination of the long-term behaviours of its
successors.

The default candidates for long-term behavior that we have alluded to
are of course the \emph{stationary distributions}, the fixpoints of
$\klex{f}$.  That our ca-homomorphisms are a generalization, is
apparent from the Kleisli-coinductive solutions: the matrix version by
Lemma~\ref{lemma:markov-matrix} of the coinduction property gives the
linear fixpoint equation $E^\dag = E^\dag E$, such that the rows of
$E^\dag$ are logically independent unitary left eigenvectors of $E$,
which is the standard characterisation of stationary distributions.
Note that, even though the intuition of distributions represented as
formal convex combinations of states is quite dominant, the target
space $Y$ is completely abstract in this framework; any convex set
will do from an axiomatic perspective.  The classical presentation can
be retrieved as the distinguished target $\mathbb T$-algebra $(TX,
\mu_X)$.

Regarding the degrees of freedom in ca-homomorphisms, Markov chains
are an interesting escalation over exact dynamical systems: On the one
hand, they support an analogous concept of coproduct structure, namely
the so-called \emph{communicating classes}, which are the strongly
connected components of the non-zero transition probability graph.  On
the other hand, they have a well-known additional degree of freedom:
even irreducible (totally connected) Markov chains can have ambiguous
stationary distribution, if they are \emph{periodic}; that is, some
state has zero transition probability to itself in $n$ steps, for
arbitrarily large $n$.

Note that, for subtle interpretative reasons, the direction of time in
this subsection is reversed with respect to the previous one; the
language of equation coalgebras speaks about the future rather than
the past, whereas the algebra constructs the past rather than the
future; in philosophical terms, the latter is governed by \emph{final}
rather than \emph{causal} reasons.

\section{Applications of Co-Kleisli Induction}
\label{coapp}

\subsection{Economic Games}

As the first application of co-Kleisli induction, we review the
coalgebraic presentation of economic game theory from
\cite{Abramsky2012}.  They construct perfect information games in
extensional form by means of anamorphisms into the final coalgebra of
a suitable game functor.  We adapt their formalization slightly.

\begin{definition}[Game Functor]
  Fix a set $\mathcal{A}$ of \emph{agents} and a set $\mathcal{C}$ of
  \emph{choices}.  The game functor is defined as follows:
  \begin{equation*}
    GX = \mathbb{R}^{\mathcal{A}} + \mathcal{A} \times X^{\mathcal{C}}
  \end{equation*}
  Elements are of either of the forms $\iota_1(u : \mathcal{A} \to
  \mathbb{R})$ or $\iota_2(a, m : \mathcal{C} \to X)$.  They are
  understood as stating, respectively, that `the game terminates with
  payoff $u(x)$ for all agents $x \in \mathcal{A}$', or that `the game
  continues with agent $a$'s turn and continuation $m(c)$ for any
  possible choice $c \in \mathcal{C}$ that $a$ can make'.
\end{definition}

In \cite{Abramsky2012}, possibly infinite game trees are unfolded from
coalgebras $(X, e)$ that model games as transition systems.  The
concrete states from $X$ are erased by the anamorphism as usual, only
the transition structure is retained.  By contrast, we shall consider
the cofree comonad $G^\infty$, and the lifted $G^\infty$-coalgebras
$(X, e^\sharp)$.  In comparison, they can be understood as running
`with logging'; the original states are retained as node labels in the
tree.  This allows for concise treatment of game tree evaluations as
node label processors.

We shall present only a single, very well-known evaluation algebra, in
order to demonstrate the applicability of our approach.  The field of
game theory is broad and deep, and in great need of formal frameworks
that give concise and elegant notations.  Thus we mark the extension
of the present sketch as an interesting topic for future research,
which however requires a great deal more effort and space.

One of the historically and logically most basic evaluation techniques
for games is \emph{backward induction} \cite{vonNeumann1944}.

The typical application of backward induction, especially in economic
uses of game theory such as discussion in \cite{Abramsky2012}, is the
computation of expected payoff, under the assumption that all agents
choose such as to maximize their own payoff.  This notion can be
stated extremely concisely as the generating $G$-algebra
$(\mathbb{R}^{\mathcal{A}}, k_0)$ of a cobasic $G^\infty$-algebra
$(\mathbb{R}^{\mathcal{A}}, k = k_0 \circ
\chi_{\mathbb{R}^{\mathcal{A}}})$:
\begin{align*}
  k_0\bigl(\iota_1(u)\bigr) &= u & k_0\bigl(\iota_2(a, m)\bigr) &= m\bigl(\operatorname{arg\,max}
  c\mathpunct. m(c)(a)\bigr)
\end{align*}

It follows that any co-Kleisli-inductive solution $k^\ddag : G^\infty
X \to \mathbb{R}^{\mathcal{A}}$ satisfies $k^\ddag(g) = u$ if
$\theta_X(g) = \iota_1(u)$, and the fairly convoluted $(k^\ddag \circ
m)\bigl(\operatorname{arg\,max} c\mathpunct. (k^\ddag \circ
m)(c)(a)\bigr)$ if $\theta_X(g) = \iota_2(a, m)$, which is on second
thought a natural formalization of the above prose description for
game trees in general, regardless of the game being played.  The
canonical extension to a ca-homomorphism from a concrete game
coalgebra $(X, f)$, namely $k^\ddag \circ f$, restricts the evaluation
to legal trees with a given starting state.

It is well-known and appreciated in game theory that there is a unique
solution, if only finite games are considered.  In our framework, that
is to say that finite games can be accomodated by the \emph{recursive
  subcomonad} of $G^\infty$~\cite{Uustalu2011135}.  We conjecture that
some standard techniques for forcing canonical solutions in infinite
cases, such as \emph{discounting}, can be expressed conveniently in
terms of a non-cobasic evaluation algebra.

As a sidenote, the very term \emph{backward induction} embodies the
characteristic difficulty of writing about scientific modelling,
namely the required double reading with regard to formal
representation and phenomenological interpretation, respectively.  As
we have noted in the preceding section, the direction of time, and
hence the word `backward' is a concept of the phenomenological
dimension; by contrast the algebraic structure and hence the word
`induction' is a concept of the formal dimension, making the composite
term essentially a category mistake.

\subsection{Lindenmayer Fractals}

The following, last example application, while another instance of
co-Kleisli-induction for a fixed algebra, has added value in two
respects: Firstly, it demonstrates the definition of a domain-specific
comonad as part of the formal modelling framework.  Secondly and
paradoxically, it makes perfectly reasonable use of essentially
non-well-founded induction.

It is a well-known statement, and frequent subject of mathematical
instruction, that certain fractal shapes can be specified by
Lindenmayer systems (L-systems), a formalism akin to Chomsky grammars
that has originally been developed for formal modelling of the growth
behavior of
plants~\cite{Lindenmayer1968,Prusinkiewicz1990,Rozenberg1980}.  This
statement, while morally true, is to be taken with a pinch of salt;
considerable effort is required for a satisfactorily rigorous
explication.  We shall demonstrate that the framework of
ca-homomorphisms is a natural and expressive background for this line
of reasoning.  Confer \cite{Hasuo2010} for another account of fractals
that builds on ca-homomorphisms (their diagram~3), but proceeds in an
interestingly different way.

\begin{definition}[Curve]
  Fix some real Banach space $V$, say without loss of
  generality $V = \mathbb{R}^2$.  Fix a real interval $I$.  An
  $I$-curve is a continuous map $f : I \to V$.
\end{definition}

\begin{definition}[Unit Curve]
  Fix some unit vector $\vec e \in V$ with $\vec e \neq \vec 0$, say
  without loss of generality $\vec e = (1, 0)$.  An $\vec e$-unit
  curve is a $[0, 1]$-curve $f$ with $f(0) = \vec 0$ and $f(1) = \vec
  e$.  We write $C_1$ for the set of all unit curves.
\end{definition}

\begin{example}
  The simplest unit curve is the \emph{stroke}: $\mathit{str}(x) = x
  \vec e$.
\end{example}

\begin{definition}[Context-Free L-System]
  As observed in \cite{TranconyWidemann2012b}, simple context-free
  L-systems, with nonterminal symbols only, are finite coalgebras of
  the (nonempty) list functor $\Lind[0]X = X^+$: The carrier is the
  set of nonterminals, and the operation is simply the set of
  production rules.  The traditional semantics of an L-system $(X, f)$
  is the sequence of iterations $(\klex[,n]{f}(s))$, seeded with a
  start symbol $s \in X$.
\end{definition}

We write $\nil$ for the empty list, and $x \cons r$ for a list with
first element $x$ and rest list $r$, as in the programming language
ML.  Complex lists $x_1 \cons \cdots \cons x_n \cons \nil$ are
condensed to $\llist x_1, \dots, x_n \rlist$.  The operator $\oplus$
denotes binary concatenation of lists.

\begin{definition}[Turn]
  Fix some group $(G, {\cdot}, i)$ of linear isometries on $V$, here
  the special orthogonal group of rotations $\rho_\alpha= \left(\begin{smallmatrix}
        \cos \alpha & -\!\sin \alpha
        \\
        \sin \alpha & \hphantom{+}\!\cos \alpha
      \end{smallmatrix}\right)$
  parameterized by angle $\alpha$.
  The elements of $G$ are called \emph{turns}.  The action on vectors
  is written ${\bullet} : G \times V \to V$.
\end{definition}

\begin{definition}[Fractal L-System]
  L-systems for specifying fractals are extensions of simple
  context-free L-systems, that is, finite coalgebras of a more complex
  functor.  In particular, they come with an additional real-valued
  shrink factor as a per-rule attribute, as well as terminal symbols
  for turns.  As hinted in the short paper
  \cite{TranconyWidemann2012b}, these can be expressed conveniently in
  terms of composition $\Lind[fr] = \mathcal{S} \Lind[0] \mathcal{T}$
  with additional functors $\mathcal{S} = \mathbb{R} \times
  \mathrm{Id}$ and $\mathcal{T}= G + \mathrm{Id}$, respectively.
  Note that $\Lind[fr]$ is monotonic and even preserves inclusions.
\end{definition}

\begin{example}[Koch Curve]
  The following singleton L-system specifies a fractal shape known as
  the Koch curve.  We use a visually evocative, short notation where
  the shrink factor appears as superscript to the production rule
  arrow, and nonterminals and turns, represented by their angles, are
  simply juxtaposed.
  \begin{equation*}
    K \stackrel3\longrightarrow K\, (+\alpha)\, K\, (-2\alpha)\, K\, (+\alpha)\, K
  \end{equation*}
  This contains the same information as the more formal, but also much
  more cumbersome full notation of $f : \{ K \} \to \Lind[fr] \{ K \}$
  as
  \begin{equation*}
    f(K) = \bigl(3, \bigl\llist \iota_2(K), \iota_1(r_{+\alpha}),
    \iota_2(K), \iota_1(r_{-2\alpha}), \iota_2(K), \iota_1(r_{+\alpha}),
    \iota_2(K) \bigr\rlist\bigr)
  \end{equation*}
\end{example}

In order for an L-system to meaningfully specify a fractal curve,
additional constraints must be met.  These are not readily expressed
in syntactic terms, and because of the inherently self-referential
nature of L-systems, great care must be exercised not to state them in
a circular way.

\begin{definition}[Well-Formed Rule]
  \label{def:wfr}%
  We give an abstract geometric interpretation of expressions of type
  $\Lind[fr]X$, under the assumption that each nonterminal represents
  a unit curve.  To this end, we track the overall change of position
  and orientation effected by sequentially tracing curves and
  executing turns, as in `turtle' graphics, disregarding the shrink
  factor.
  \begin{align*}
    \mathit{span} &: \Lind[0]\mathcal{T}X \to V &
    \mathit{dir} &: \Lind[0]\mathcal{T}X \to G
    \\
    \mathit{span}(t) &=
    \begin{cases}
      \vec 0 & \text{if}\quad t = \nil
      \\
      \rho \bullet \mathit{span}(u) & \text{if}\quad t = \iota_1(\rho) \cons u
      \\
      \vec e + \mathit{span}(u) & \text{if}\quad t = \iota_2(x) \cons u
    \end{cases}
    &
    \mathit{dir}(t) &=
    \begin{cases}
      i & \text{if}\quad t = \nil
      \\
      \rho \cdot \mathit{dir}(u) & \text{if}\quad t = \iota_1(\rho) \cons u
      \\
      \mathit{dir}(u) & \text{if}\quad t = \iota_2(x) \cons u
    \end{cases}
  \end{align*}
  A rule is called \emph{well-formed} if and only if it is properly
  shrinking, spans the unit vector when shrunk, and has balanced
  turns: $\mathit{wfr}\bigl((a, t)\bigr) \iff a > 1 \land
  \mathit{span}(t) = a \vec e \land \mathit{dir}(t) = i$.
\end{definition}

\begin{lemma}
  \label{lemma:span-dir-abs}%
  The abstract interpretation is additive.
  \begin{align*}
    \mathit{span}(t \oplus u) &= \mathit{span}(t) + \mathit{dir}(t)
    \bullet \mathit{span}(u)
    &
    \mathit{dir}(t \oplus u) &= \mathit{dir}(t) \cdot \mathit{dir}(u)
  \end{align*}
\end{lemma}

\begin{lemma}
  \label{lemma:abs}%
  Well-formedness is additive:
  $\mathit{wfr}\bigl((a, t)\bigr) \land \mathit{wfr}\bigl((b, u)\bigr) \implies \mathit{wfr}\bigl((a + b, t \oplus u)\bigr)$.
\end{lemma}

\begin{lemma}
  Well-formedness is invariant under the fractal Lindenmayer
  functor.
\end{lemma}

This result allows us to abstract from the contents of a rule,
retaining only the shape, and to construct a subfunctor for
well-formed rules.

\begin{definition}[Rule Shape]
  Let $1 = \{ * \}$ be the final object, with unique morphisms $!_X :
  X \to 1$.  We call $\Lind[fr]1$ the \emph{shape space}, and
  $\Lind[fr]\,!_X : \Lind[fr]X \to \Lind[fr]1$ the \emph{shape map}.
\end{definition}

\begin{lemma}
  A rule is well-formed if and only if its shape is.
\end{lemma}

\begin{definition}[Well-Formed Fractal L-System]
  We obtain a subfunctor $\Lind[wfr] = \Lind[fr]
  \rvert_{\mathit{wfr}}$ of the general fractal Lindenmayer functor by
  restriction to well-formed rules.
\end{definition}

\begin{definition}[Rule Interpretation]
  The concrete interpretation of rules differs from the abstract
  interpretation with $\mathit{span}$ by actually composing the
  subcurves listed in a rule, concatenating their domains.  We model
  this idea as a $\Lind[fr]$-algebra $(Y, \mathit{draw})$ over the
  general function space $Y = V^{[0, 1]}$, not assuming continuity for
  now.

  In order to obtain the shape of a unit curve, both the domain and
  the result need to be compressed, according to the length of the
  rule and the shrink factor, respectively.  In the following, let
  $\ell$ denote the length of list $t \in \Lind[0]\mathcal{T}Y$.
  \begin{align*}
    \mathit{draw}\bigl((a, t)\bigr)(z) &= a^{-1} \mathit{step}(t)(\ell z)
    \qquad \text{where}
    \\
    \mathit{step}(t)(x) &=
    \left\{\begin{array}{@{}l@{\quad\text{if}\quad}ll@{\qquad}c}
      \vec 0 & z \in [0, 1] & t = \iota_1(\rho) \cons u & \text{(A)}
      \\
      \vec 0 & z = 0 & t = \iota_2(f) \cons u & \text{(B)}
      \\
      f(x) & z \in (0, 1) & t = \iota_2(f) \cons u & \text{(C)}
      \\
      \vec e & z = 1 & t = \iota_2(f) \cons u & \text{(D)}
      \\
      \rho \bullet \mathit{step}(u)(z - 1) & z > 1 & t = \iota_1(\rho) \cons u & \text{(E)}
      \\
      \vec e + \mathit{step}(u)(z - 1) & z > 1 & t = \iota_2(f) \cons u & \text{(F)}
    \end{array}\right.
  \end{align*}
  The auxiliary function $\mathit{step}$ has dependent type; for every
  argument $t$ of length $\ell > 0$, it yields a function of type
  $V^{[0, \ell]}$.  Note that the case $t = \nil$ does not
  occur, even recursively.
\end{definition}

By overlapping the subcurves at their glueing points in the
composition, and even sending turns to constant points, we depart
radically from the way of \cite{Hasuo2010}, where \emph{injectivity}
is a major theoretical goal, and successfully solved technical
obstacle.  Whereas they have been concerned with faithful
\emph{representation} of fractal space, our approach is more a
\emph{computational} account: By virtue of the piecewise constancy
introduced by turns, a point $f(x)$ for random $x$ can be computed
\emph{exactly} by finite recursion \emph{almost surely}, except for
some pathological L-systems.  We discuss this property more deeply in
a forthcoming companion paper.

\begin{lemma}
  \label{lemma:step-span}%
  The endpoint of $\mathit{step}$ is predicted by $\mathit{span}$:
  $\mathit{step}(t)(\ell) = \mathit{span}(t)$.
\end{lemma}

\begin{theorem}
  \label{theorem:wellf-unit}%
  Interpretation of well-formed rules preserves unit curves:
  $\mathit{draw}(r) \in C_1$ for all $r \in
  \mathcal{L}_{\mathrm{wfr}}(C_1)$.
\end{theorem}

\begin{proof}[Proof (Sketch)]
  By piecewise continuity with agreement at glueing points.
\end{proof}

\begin{definition}[Well-Formed Rule Interpretation]
  We obtain an $\mathcal{L}_{\mathrm{wfr}}$-algebra $(C_1,
  \mathit{draw}_{\mathrm{wfr}})$ by restricting rule interpretation
  $\mathit{draw}$ to unit curves.
\end{definition}

\begin{definition}
  Let $\alpha : \Lind[(w)fr] \Rightarrow \mathcal P$ be the obvious
  natural transformation that extracts the set of nonterminal symbols
  occurring in a rule.  Let $\mathbb P = (\mathcal P_\omega, \eta,
  \mu)$ be the well-known \emph{finite} powerset monad.  Now consider
  the cofree comonad over $\Lind[wfr]$.  The object $\Lind[wfr]^\infty
  X$ can be considered as the space of non-well-founded $X$-labelled
  trees of nested well-formed rules.  From these we can extract sets
  of nestedly occurring rule shapes: The natural transformation
  \begin{itemize}
  \item $\rho = \eta \circ \chi : \Lind[wfr]^\infty \Rightarrow
    \mathcal P_\omega \Lind[wfr]$ extracts the single root shape;
  \item $\sigma = \alpha F^\infty \circ \theta : \Lind[wfr]^\infty
    \Rightarrow \mathcal P_\omega \Lind[wfr]^\infty$ extracts the
    immediate subtrees.
  \end{itemize}
  Thus the extraction of all shapes occurring at some finite depth in
  $t \in \Lind[wfr]^\infty X$ is defined as $S_X(t) = \bigcup_{n =
    0}^\infty (\mathcal P_\omega !_X \circ \klex{\rho_X} \circ
  \klex[,n]{\sigma_X})(t)$, where Kleisli extension is over $\mathbb
  P$.  Confer the relation $(\leadsto_e)$ from
  Definition~\ref{def:dyn-rels}.
\end{definition}

\begin{definition}
  We obtain the \emph{shape-finitary} subcomonad $\mathbb D$ of the
  cofree comonad of well-formed L-systems: $DX = \{ t \in
  \Lind[wfr]^\infty X \mid S_X(t) ~\text{finite} \}$.  Clearly, all
  actual such L-systems $(X, f)$, which are \emph{finite}
  $\Lind[wfr]$-coalgebras, extend to $\mathbb D$-coalgebras $(X,
  f^\sharp)$.  Note the analogy to \emph{rational coalgebras}
  \cite{Adamek2006,TranconyWidemann2005}.

  Finally we construct an interpretation (coequation) algebra, the
  very goal of this whole subsection, by extending
  $\mathit{draw}_{\mathrm{wfr}}$ to a cobasic
  $\Lind[wfr]^\infty$-algebra, and restricting its domain to
  shape-finitary trees: $k = \mathit{draw}_{\mathrm{wfr}} \circ
  \chi_{C_1}\rvert_{DC_1}$.
\end{definition}

The ca-homomorphisms $h : (X, f) \to (C_1, k)$ map nonterminal symbols
of L-systems to unit curves, with the obvious self-similar geometric
consistency condition.  Now we can make the point of our nontrivial
efforts and restrictions.

\begin{theorem}
  \label{theorem:conn-rec}%
  The coequation $D$-algebra $(C_1, k)$ is corecursive; a unique
  ca-homomorphism exists from any $\mathbb D$-coalgebra.
\end{theorem}

\begin{proof}[Proof (Sketch)]
  By uniform continuity and uniqueness on the dense subdomain of
  recursive glueing points.
\end{proof}

Note that, while the comonadic structure appears not to be used by the
cobasic algebra, it is required for expressing the restriction to
shape-finitary L-systems, which in turn is a necessary condition for
our proof; infinitary counterexamples can be conceived.



\appendix



\bibliography{arxived}


\clearpage
\section{Omitted Proofs}

This section gives or completes all proofs that we consider
informative or non-elementary.  Some boring immediate or inductive
proofs are omitted altogether.

\begin{proof}[Proof of Lemma~\ref{lemma:alg-tup}]
  Let $F$ preserve coproducts.  Let index $i$ range over some set $I$
  in all of the following.  Then, for any family of objects $(X_i)$,
  there is a unique isomorphism $j : \coprod_i F X_i \to F \coprod_i
  X_i$ such that, for any family of morphisms $(h_i : X_i \to Y)$, we
  have $[F h_i] = F [h_i] \circ j$.  Now let $(X_i, f_i)$ be a family
  of $F$-coalgebras.  Then $(X, f) = (\coprod_i X_i, j \circ \coprod
  f_i)$ is again an $F$-coalgebra.  Let $(Y, g)$ be a fixed
  $F$-algebra.  The ca-homomorphisms of type $h : (X, f) \to (Y, g)$
  are in one-to-one correspondence with the families of morphisms
  $(h_i)$ where all members are ca-homomorphisms $h_i : (X_i, f_i) \to
  (Y, g)$, namely $h = [h_i]$:
  \begin{equation*}
    \xymatrix{
      \coprod_i X_i \ar[dd]_{[h_i]} \ar[rrr]^{\coprod_i f_i} & & & \coprod_i F X_i \ar[dd]^{[F h_i]} \ar[r]^{j} & F \coprod_i X_i \ar@/^4ex/[ldd]^{F [h_i]} &
      \\
      & X_i \ar[ld]^{h_i} \ar[lu]_{\iota_i} \ar[r]^{f_i} & F X_i \ar[ru]^{\iota_i} \ar[rd]_{F h_i}
      \\
      Y & & & F Y \ar[lll]^{g}
    }
  \end{equation*}
  For products of algebras dually.
\end{proof}

\begin{proof}[Proof of Lemma~\ref{lemma:tau}\,\&\,\ref{lemma:theta}]
  By natural laws of coproduct and monad, and catamorphic $\mu$.
  \begin{equation*}
    \xymatrix{
      FF^* \ar@{=>}[d]_{F\eta F^*} \ar@{=}[rrd] \ar@{=>}@/_14ex/[ddd]_{\kappa F^*}
      \\
      FF^*F^* \ar@{=>}[d]^{\iota2} \ar@{=>}@/_7ex/[dd]_{\tau F^*} \ar@{=>}[rr]^{F\mu} && FF^* \ar@{=>}[d]_{\iota_2} \ar@{=>}@/^7ex/[dd]^{\tau}
      \\
      F^* {+} FF^*F^* \ar@{=>}[d]^{\mathrm{in}F^*} \ar@{=>}[rr]^{F^* + F\mu} && F^* {+} FF^* \ar@{=>}[d]_{[\mathrm{id}F^*, \tau]}
      \\
      F^*F^* \ar@{=>}[rr]_{\mu} && F^*
    }
  \end{equation*}
  For $\theta$ dually.
\end{proof}

\begin{proof}[Proof of Lemma~\ref{lemma:mu}\,\&\,\ref{lemma:nu}]
  The former clause by compatibility of $f$ with $\mu$; the latter
  clause by naturality of $\mu$.
  \begin{align*}
    \xymatrix{
      TTX \ar[r]^{\mu_X} \ar[d]_{Tf} & TX \ar[d]^{f}
      \\
      TX \ar[r]_{f} & X
    } &&
    \xymatrix {
      TTX \ar[r]^{\mu_X} \ar[d]_{TTh} & TX \ar[d]^{Th}
      \\
      TTY \ar[r]_{\mu_Y} & TY
    }
  \end{align*}
  For $\nu$ dually.
\end{proof}

\begin{proof}[Proof of Lemma~\ref{lemma:dag}\,\&\,\ref{lemma:codag}]
  Proposition laid out in underlying category.
  \begin{equation*}
    \xymatrix{
      TX \ar[d]_{Te^\dag} \ar[rd]|{\klex{(e^\dag)}} & X \ar[l]_{e} \ar[d]^{e^\dag}
      \\
      TTY \ar[r]_{\mu_Y} & TY
    }
  \end{equation*}
  For $k^\ddag$ dually.
\end{proof}

\begin{proof}[Proof of Lemma~\ref{lemma:dag-subst}\,\&\,\ref{lemma:codag-subst}]
  By composition of Lemma~\ref{lemma:mu}, second clause, with
  Lemma~\ref{lemma:dag}.
  \begin{equation*}
    \xymatrix{
      TX \ar[d]_{Te^\dag} \ar[rd]|{\klex{(e^\dag)}} & X \ar[l]_{e} \ar[d]^{e^\dag}
      \\
      TTY \ar[r]^{\mu_Y} \ar[d]_{TTh} & TY \ar[d]^{Th}
      \\
      TTZ \ar[r]_{\mu_Z} & TZ
    }
  \end{equation*}
  For $k^\ddag$ dually.
\end{proof}

\begin{proof}[Proof of Theorem~\ref{theorem:dag}\,\&\,\ref{theorem:codag}]
  By composition of Lemma~\ref{lemma:mu}, first clause, with
  Lemma~\ref{lemma:dag}.
  \begin{equation*}
    \xymatrix{
      TX \ar[d]_{Te^\dag} \ar[rd]|{\klex{(e^\dag)}} & X \ar[l]_{e} \ar[d]^{e^\dag}
      \\
      TTY \ar[r]^{\mu_Y} \ar[d]_{Tf} & TY \ar[d]^{f}
      \\
      TY \ar[r]_{f} & Y
    }
  \end{equation*}
  For $k^\ddag$ dually.
\end{proof}

\begin{proof}[Proof of Lemma~\ref{lemma:basic}\,\&\,\ref{lemma:cobasic}]
  By naturality of $\kappa$.
  \begin{equation*}
    \xymatrix{
      FX \ar[r]^{\kappa_X} \ar[d]_{Fh} & F^*X \ar[d]|{F^*h} & X \ar@/_4ex/[ll]_{e_0} \ar[d]^{h} \ar[l]_{e}
      \\
      FY \ar@/_4ex/[rr]_{f^\flat} \ar[r]_{\kappa_Y} & F^*Y \ar[r]_{f} & Y
    }
  \end{equation*}
  For $\chi$ dually.
\end{proof}

\begin{proof}[Proof of Lemma~\ref{lemma:dyn-cons}]
  Assume that $x \leadsto_e (t, x')$ and $x \leadsto_e (u, x')$.  Then
  there are $m, n$ such that $\klex[,m]{e}(x) = (t, x')$ and
  $\klex[,n]{e}(x) = (u, x')$.  Assume without loss of generality
  $m < n$.  Then
  \begin{equation*}
    \begin{aligned}
      (u, x') &= \klex[,n]{e}(x) \\
      &= ((\klex{e})^{n-m} \circ \klex[,m]{e})(x)
      \\
      &= (\klex{e})^{n-m}\bigl((t, x')\bigr)
      \\
      &= \klex{(\klex[,n-m]{e})}\bigl((t, x')\bigr)
      \\
      &= (t + v, x') & \text{where}~ \klex[,n-m]{e}(x') &= (v, x')
      \\
      && x' &\leadsto_e (v, x')
    \end{aligned}
  \end{equation*}
  With $v = 0$ by consistency, we obtain $t = u$.
\end{proof}

\begin{proof}[Proof of Theorem~\ref{theorem:dyn-wellf}]
  By well-founded induction: Assume $e^\dag(x')$ has been chosen for
  all $x' \prec_e x$.  Then either $x$ is minimal, and $e^\dag(x)$ can
  be chosen arbitrarily; or otherwise there are some $x', x'', t$ such
  that $x' \prec_e x$, $x'' \simeq_e x$, and $e(x'') = (t, x')$.  Then
  choose $e^\dag(x'') = (t + u, y)$ where $e^\dag(x') = (u, y)$.  By
  cancellation and Lemma~\ref{lemma:dyn-stack}, necessarily
  $e^\dag(x'') = e^\dag(x)$ for all $x'' \simeq_e x$.
\end{proof}

\begin{proof}[Proof of Theorem~\ref{theorem:dyn-group}]
  Consider the relation $\succ_e$ restricted to each equivalence class
  $C$ of $\sim_e$ in turn, by Lemma~\ref{lemma:dyn-coprod}.  Either it
  is well-founded, such that Theorem~\ref{theorem:dyn-wellf} applies
  directly; or otherwise the choice of any element $x$ partitions $C$
  into the down-set $D$ of $x$, which is an infinitely descending
  chain, and its complement $C \setminus D$, which is well-founded.
  Then fix $u, y$ arbitrarily, and choose $e^\dag(x') = (u - t,
  y)$ where by Lemma~\ref{lemma:dyn-cons}, $t$ is uniquely determined
  by $x \leadsto_e (t, x')$, for each $x' \in D$.  Proceed on $C
  \setminus D$ as in Theorem~\ref{theorem:dyn-wellf}.
\end{proof}

\begin{proof}[Proof of Theorem~\ref{theorem:wellf-unit}]
  Consider a well-formed rule $r = (a, t)$.

  The map $\mathit{step}(t)$ is defined piecewise: glueing points
  at integer arguments are fixed, by cases (B) and (D).  Consider the
  first pair of adjacent glueing points, at $0$ and $1$: they are
  either identical $\vec 0$, and the map is constant on the whole
  closed interval $[0, 1]$, by case (A), or they are $\vec 0$ and
  $\vec e$, respectively, and a unit curve is spliced continuously in
  between, by case (C).  For further pairs, continuously shifted and
  turned analogs hold, by cases (E) and (F), respectively.  Hence the
  map is continuous as a whole.

  Furthermore, we have $\mathit{step}(t)(0) = \vec 0$ by case (B), and
  $\mathit{step}(t)(\ell) = a \vec e$ by well-formedness and
  Lemma~\ref{lemma:step-span}.  Thus the rescaled map
  $\mathit{draw}(r)$ is a unit curve.
\end{proof}

\begin{proof}[Proof of Theorem~\ref{theorem:conn-rec}]
  By Lemma~\ref{lemma:basic}, the $D$-ca-homomorphisms $h : (X,
  f^\sharp) \to (C_1, k)$ can be understood as
  $\Lind[wfr]$-ca-homomorphisms $h : (X, f) \to (C_1,
  \mathit{draw}_{\mathrm{wfr}})$.  That is, they are recursive in the
  nonterminal slots of a rule.  Now proceed by considering the
  resulting curve pointwise, and case distinction in the workhorse
  function $\mathit{step}$:

  By the base cases (A), (B) and (D), the inductive definition of $h$
  in terms of induction step $\mathit{draw}$ is well-founded, and
  hence the resulting curve defined uniquely, on a dense subdomain $J
  \subseteq [0, 1]$; namely on isolated points of recursive interval
  subdivision by (B) and (D), and on closed subintervals by (A).  By
  the induction hypothesis, the recursively included subcurves are
  unit curves and hence continuous on $J$.  By finite amount of
  possible shapes, the extent of subcurves on $J$ is bounded, but they
  are shrunk recursively by unbounded factors.  Thus the resulting
  curve is totally bounded and hence uniformly continuous on $J$.  By
  a standard result, a uniformly continuous map on a dense subdomain
  extends uniquely to a continuous map on its closure, the whole unit
  domain $[0, 1]$.  Clearly, that extension is a unit curve.
\end{proof}

\clearpage
\section{Supplementary Figures}

This appendix illustrates the construction of fractal curves from
L-Systems.  Two example systems for very well-known fractals are
considered.

\begin{itemize}
\item The Koch curve
  \begin{equation*}
    K \stackrel{3}\longrightarrow K \, l \, K \, r \, r \, K \, l \, K
  \end{equation*}
\item The Sierpinski triangle
  \begin{align*}
    U &\stackrel{2}\longrightarrow l \, D \, r \, U \, r \, D \, l
    \\
    D &\stackrel{2}\longrightarrow r \, U \, l \, D \, l \, U \, r
  \end{align*}
  where $U$ yields the standard upright triangle, and $D$ its downward
  mirror image.
\end{itemize}

The fractals are visualized by equidistantly spaced samples from the
domain interval.  The unit vector $\vec e$ spans the baseline from
left to right.  Note that \emph{analytically exact} coordinates can be
given for each sampled point by recursive symbolic interpretation.
Sparse points are connected by dashed lines as a visual aid; these are
not valid interpolations of the fractal.  Note also that there may be
fewer visible points than samples due to non-injectivity.

The figures have been generated by programs written by an author in
Haskell and {\sffamily R}.  The source code is available on request.

\begin{figure}[h]
  \centering
  \pgfimage[width=.8\textwidth]{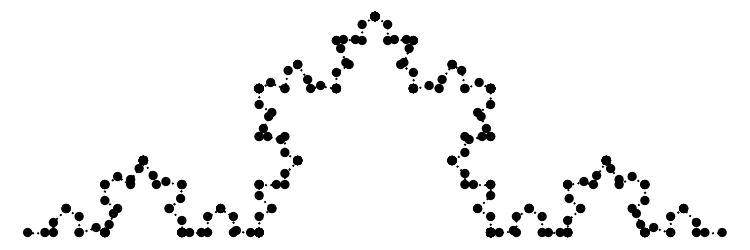}
  \caption{Koch curve $h(K)$, sampled every $1/666$}
\end{figure}

\begin{figure}[h]
  \centering
  \pgfimage[width=.8\textwidth]{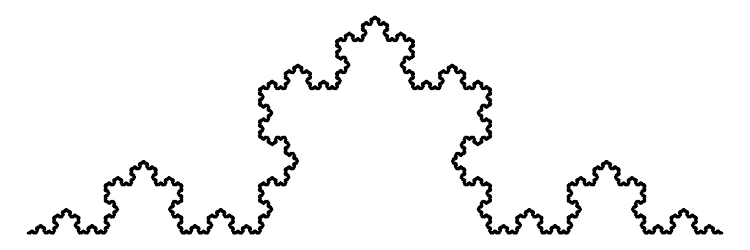}
  \caption{Koch curve $h(K)$, sampled every $1/400000$}
\end{figure}

\begin{figure}[p]
  \centering
  \pgfimage[width=.8\textwidth]{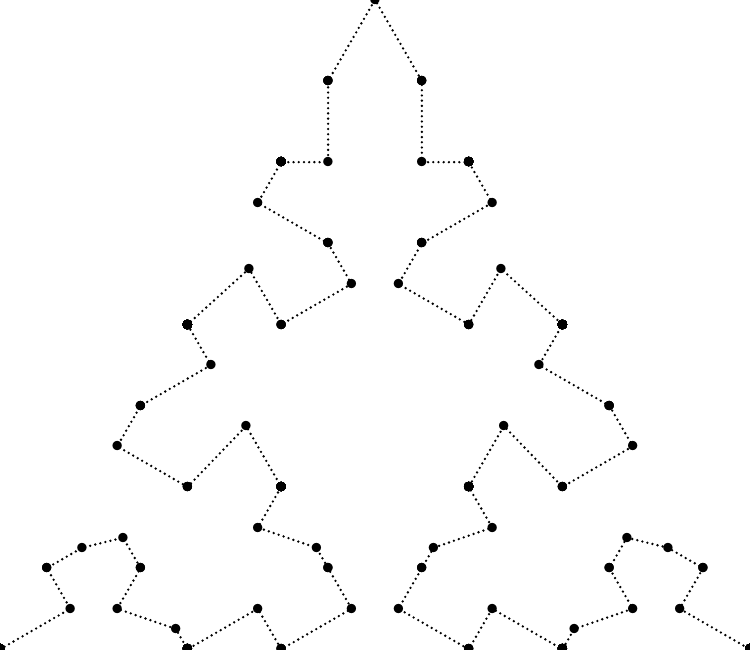}
  \caption{Sierpinski triangle $h(U)$, sampled every $1/666$}
\end{figure}

\begin{figure}[p]
  \centering
  \pgfimage[width=.8\textwidth]{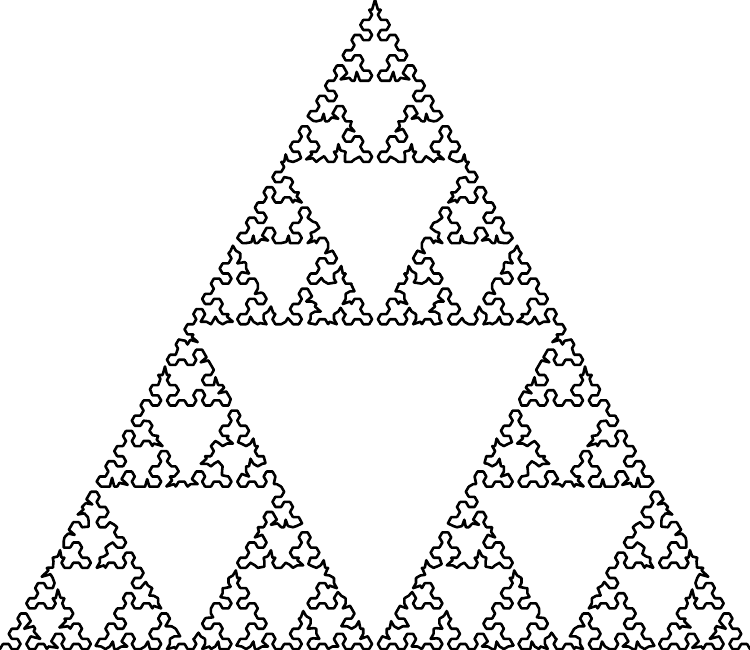}
  \caption{Sierpinski triangle $h(U)$, sampled every $1/400000$}
\end{figure}

\end{document}